\pdfoutput=1
\documentclass[conference]{IEEEtran}
\IEEEoverridecommandlockouts
\usepackage{amsmath,amssymb}
\usepackage{graphicx}
\usepackage{booktabs}
\usepackage{array}
\usepackage{multirow}
\usepackage{xcolor}
\usepackage{tikz}
\usetikzlibrary{arrows.meta,positioning,fit,calc,decorations.pathreplacing}
\usepackage{pgfplots}
\pgfplotsset{compat=1.17}
\usepackage{url}
\usepackage{cite}

\begin{document}

\title{SkipLane: Write-Once, Conflict-Free Accumulation\\for Zero-Skipping CNN Engines on FPGAs\thanks{This work has been submitted to the IEEE for possible publication. Copyright may be transferred without notice, after which this version may no longer be accessible.}}

\author{\IEEEauthorblockN{Aditya Saboo and Pratibha Verma}
\IEEEauthorblockA{Department of Electrical Engineering, Indian Institute of Technology Indore, Indore, India \\
pratibhaverma@iiti.ac.in}}

\maketitle

\begin{abstract}
Skipping zero activations saves multiplications but makes accumulation harder: products from a compressed input window belong to different output columns. In our earlier engine, merging them by read-modify-write (RMW) of a row buffer made sparse inputs slower than dense inputs, even at 87\% sparsity. Prior designs merge kernel rows by RMW, or reduce bank conflicts with extra hardware without removing them. SkipLane makes the whole accumulation path write-once: each kernel row has its own lane, which retires up to three finished columns per cycle into banks that cannot conflict, and lanes and input channels are added only at readout. Multipliers idled by zero weights compute a second filter. In RTL simulation, the engine stays within 0.5\% of its frontend bound up to 70\% sparsity and is 1.10--1.26$\times$ faster than an SCNN-style scatter accumulator with 18 banks and ideal RMW at 50--80\% sparsity. Six engines meet 140\,MHz on a PYNQ-Z2 in 21.1k LUTs and run LeNet-5 bit-exact on the board: conv1 needs 1.94$\times$ fewer engine cycles than a dense engine (1.26$\times$ end to end), while conv2, with 13-pixel rows, is 0.87$\times$ as fast.
\end{abstract}


\section{Introduction}
After ReLU, 50--85\% of the activations in a CNN layer are often zero \cite{sparsepe}, and pruning makes many weights zero as well \cite{han}. Multiplying by zero is wasted work, so many accelerators skip it \cite{cnvlutin,scnn,sparten,eyerissv2,sparsepe,zasca,nullhop}.

We started from a working dense LeNet-5 accelerator on a PYNQ-Z2 board \cite{corevision} and changed it to skip zero pixels. The multipliers then had less to do, but the engine became \emph{slower}: at 50\% sparsity a full image took 2.7$\times$ longer than with dense input (Fig.~\ref{fig:speedup}). The reason is accumulation. When zeros are skipped, the products that make up one output pixel arrive at different times, and our engine merged them by RMW of a row buffer (Sec.~\ref{sec:slow}).

This paper fixes that problem. Liu et al.\ \cite{liu21} shift a kernel row's partial sums through three registers but merge kernel rows by RMW of buffers; scatter designs \cite{scnn,gaas,zhang25} reduce bank conflicts with extra hardware. Our contributions are:
\begin{itemize}
\item a \textbf{write-once accumulation path} for a streaming engine: a line-buffer window feeds all three kernel rows in the same pass, each lane retires up to three finished columns per cycle into its own conflict-free banks, and lanes and channels are added only at readout, so no partial sum is read back (Sec.~\ref{sec:arch});
\item \textbf{filter packing}, which lets the nonzero weights of two filters share one 3$\times$3 engine, so weight and activation zeros are skipped in the same pass (Sec.~\ref{sec:pack});
\item a \textbf{comparison with an SCNN-style scatter accumulator} on the same frontend, and a bit-exact LeNet-5 run on a PYNQ-Z2 board (Sec.~\ref{sec:eval}).
\end{itemize}

\section{Background and Problem}

\subsection{Platform}
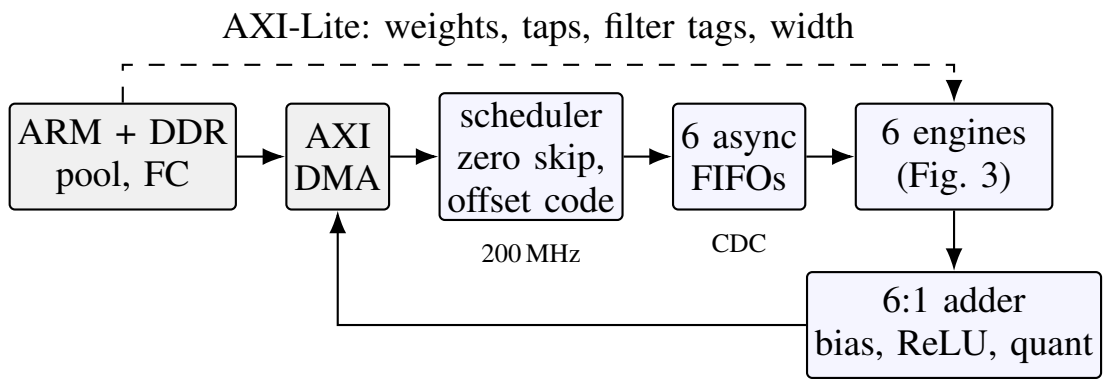
\begin{figure}[t]
\centering
\begin{tikzpicture}[font=\scriptsize, node distance=2.2mm and 3.2mm,
  b/.style={draw, rounded corners=1pt, minimum height=7mm, align=center, fill=blue!4, inner sep=1.5pt},
  p/.style={draw, rounded corners=1pt, minimum height=7mm, align=center, fill=gray!12, inner sep=1.5pt},
  arr/.style={-{Latex[length=1.5mm]}}]
\node[p] (ps) {ARM + DDR\\pool, FC};
\node[p, right=of ps] (dma) {AXI\\DMA};
\node[b, right=of dma] (sch) {scheduler\\zero skip,\\offset code};
\node[b, right=of sch] (fifo) {6 async\\FIFOs};
\node[b, right=of fifo, minimum width=13mm] (eng) {6 engines\\(Fig.~\ref{fig:arch})};
\node[b, below=4mm of eng] (l3) {6:1 adder\\bias, ReLU, quant};
\draw[arr] (ps) -- (dma);
\draw[arr] (dma) -- (sch);
\draw[arr] (sch) -- (fifo);
\draw[arr] (fifo) -- (eng);
\draw[arr] (eng) -- (l3);
\draw[arr] (l3.west) -| (dma.south);
\draw[arr, dashed] (ps.north) -- ++(0,2.5mm) -| node[pos=0.25, above]{AXI-Lite: weights, taps, filter tags, width} (eng.north);
\node[font=\tiny, below=0.3mm of sch] {200\,MHz};
\node[font=\tiny, below=0.3mm of fifo] {CDC};
\end{tikzpicture}
\caption{Overall flow. DMA sends six input channels to the scheduler, which drops zeros and adds offsets. Asynchronous FIFOs move the records into the 140\,MHz engine clock. The six engine outputs are summed, biased, passed through ReLU, requantized and sent back.}
\label{fig:flow}
\end{figure}

The accelerator runs on the Xilinx xc7z020 (PYNQ-Z2). The ARM core configures the accelerator over AXI4-Lite, sends feature maps by AXI DMA, and runs pooling and the fully connected layers. The programmable logic runs the convolutions with six 3$\times$3 engines (54 DSP multipliers) and a 6:1 channel adder with bias, ReLU and requantization (Fig.~\ref{fig:flow}). We use a LeNet-5 network \cite{lenet} in which both 5$\times$5 convolutions are replaced by 3$\times$3 ones (conv1: 1$\rightarrow$6 channels on 28$\times$28 inputs; conv2: 6$\rightarrow$16 channels on 13$\times$13 inputs), trained on MNIST, 8-bit.

\subsection{Compressed row stream}
A scheduler reads each input row and drops zero pixels, except the first three: the first two records of a row meet only $w_0$, and $w_0,w_1$, which is correct only for the pixels at columns 0 and 1; later records meet all three weights. Each kept pixel is sent as a record $\{last, z, x\}$: the pixel value $x$, the number $z$ of zeros skipped so far in the row (the \emph{offset}), and a flag for the last record. Record $m$ therefore belongs to column $c_m = m + z_m$ (Fig.~\ref{fig:example}b). Each engine stores two compressed rows in line buffers and receives the third as it streams in, so its three kernel rows work on their three input rows in parallel, not one after another as in \cite{liu21}. This K$-$1 line-buffer window follows AMD's Vitis HLS example \cite{ug1399}, built in RTL on compressed rows. In every cycle the engine reads a 3$\times$3 window: three consecutive records from each of the three rows. Each row then moves forward by one \emph{record}, not one column. As zeros are removed before the engine, a window always holds the next kept pixels; Sparse-PE \cite{sparsepe} instead searches $n$ chunks ahead, with a selector that grows with $n$; we need no selector but scan every pixel, on a faster clock. We number each window by its first record: window $w$ holds records $w$, $w{+}1$ and $w{+}2$ of every row, at positions $j=0,1,2$. With dense input, record $m$ is at column $m$ and this is the usual sliding window. With sparse input, columns of the three records need not be adjacent and differ between rows: in Fig.~\ref{fig:example}, window 0 covers columns 0, 1 and 2, but window 3 covers columns 6, 7 and 9, which tap alignment handles (Sec.~\ref{sec:arch}).

\subsection{Scatter to output columns}\label{sec:scatter}
Each PE holds one weight of a kernel row, $w_0$, $w_1$ or $w_2$; its index $j\in\{0,1,2\}$ is the PE's \emph{tap}. Output $o_a$ uses the pixels at columns $a$, $a{+}1$ and $a{+}2$: $o_a = x_a w_0 + x_{a+1} w_1 + x_{a+2} w_2$. So the tap alone tells where a product goes: a pixel at column $c$ multiplied by $w_j$ belongs to output column
\begin{equation}
a = c - j .
\label{eq:addr}
\end{equation}
In window $w$, the PE with tap $j$ multiplies record $w{+}j$, at column $c_{w+j}$, so its product goes to $a = c_{w+j} - j$.

\textbf{Vector engine.} We use 9 independent PEs rather than a systolic array or a row of MACs fed one broadcast pixel \cite{liu21}: each product carries its own column, so its arrival time does not affect correctness, compressed windows can be fed directly, and each PE selects its own weight, pixel and filter tag, which enables packing (Sec.~\ref{sec:pack}). The cost: products of one output, made in different windows, must be merged later (Sec.~\ref{sec:arch}). Unlike the MAC chains of \cite{liu21}, a PE freed by a zero weight can compute a second filter (Sec.~\ref{sec:pack}).

\subsection{Why the earlier engine was slow}\label{sec:slow}
Our earlier sparse engine merged equal columns within a window, put the results in a queue, and updated a row buffer with a 3-cycle RMW per partial sum. Skipping zeros reduces the number of windows, but splits each output into 2.7--2.9 partial sums (Table~\ref{tab:ablation}), so the memory updates, not the multipliers, set the run time.

\section{Key Observation}
\label{sec:prop}
Consider one kernel row (a \emph{lane}). A pixel at column $c$ contributes to at most three output columns: $c{-}2$, $c{-}1$ and $c$ (\textbf{bounded reach}). Retained pixels arrive in increasing column order, even when zeros are skipped. Once the next pixel arrives at column $c'$, no later pixel can contribute to an output column below $c'{-}2$ (\textbf{in-order completion}). The accumulator can therefore finish old columns as the input moves forward and keep only the three the latest pixel can still affect. Both hold for any zeros and weights, so also for two packed filters (trace: Fig.~\ref{fig:trace}).

\begin{figure}[t]
\centering
\begin{tikzpicture}[font=\scriptsize, x=0.7cm, y=1cm,
  kept/.style={draw=blue!60!black, fill=blue!10, minimum width=0.6cm, minimum height=0.48cm, inner sep=0pt},
  drop/.style={draw=gray!55, fill=gray!8, text=gray, minimum width=0.6cm, minimum height=0.48cm, inner sep=0pt},
  rec/.style={draw=blue!60!black, fill=blue!10, rounded corners=1pt, minimum width=0.98cm, minimum height=0.46cm, inner sep=0.5pt, font=\tiny},
  lbl/.style={font=\tiny, text=black!70},
  brc/.style={decorate, decoration={brace, mirror, amplitude=2.5pt}, thin, black!70}]
\foreach \c/\v/\k in {0/5/1,1/0/1,2/3/1,3/0/0,4/0/0,5/0/0,6/7/1,7/2/1,8/0/0,9/4/1} {
  \ifnum\k=1 \node[kept] (p\c) at (\c,0) {$\v$}; \else \node[drop] (p\c) at (\c,0) {$\v$}; \fi
  \node[lbl] at (\c,0.36) {\c};
}
\node[anchor=east, lbl] at (-0.45,0.36) {column $c$};
\node[anchor=east] at (-0.45,0) {(a) pixel $x$};
\draw[brc] (2.3,0.52) -- (-0.3,0.52) node[midway, above=2.5pt, lbl] {$o_0 = 5w_0{+}0w_1{+}3w_2$};
\draw[brc] (8.3,0.52) -- (5.7,0.52) node[midway, above=2.5pt, lbl] {$o_6 = 7w_0{+}2w_1{+}0w_2$};
\foreach \m/\c/\o/\x/\l in {0/0/0/5/0, 1/1/0/0/0, 2/2/0/3/0, 3/6/3/7/0, 4/7/3/2/0, 5/9/4/4/1} {
  \node[rec] (r\m) at (\m*1.55+0.6,-1.25) {$\{\l,\o,\x\}$};
  \node[lbl] at (\m*1.55+0.6,-1.58) {$m{=}\m$};
  \node[lbl, anchor=west] at ($(r\m.north east)+(-0.05,0.07)$) {};
}
\foreach \m/\c in {0/0,1/1,2/2,3/6,4/7,5/9} {
  \draw[-{Latex[length=0.9mm]}, blue!50!black, very thin] (\c,-0.27) -- (\m*1.55+0.6,-1.03);
}
\node[anchor=east] at (-0.45,-1.25) {(b) stream};
\node[anchor=east, lbl] at (-0.45,-1.5) {$\{last,z,x\}$};
\draw[brc] ($(r0.south west)+(0,-0.42)$) -- ($(r2.south east)+(0,-0.42)$) node[midway, below=2.5pt, lbl] {window 0: columns 0, 1, 2};
\draw[brc] ($(r3.south west)+(0,-0.42)$) -- ($(r5.south east)+(0,-0.42)$) node[midway, below=2.5pt, lbl] {window 3: columns 6, 7, 9};
\end{tikzpicture}

\caption{One lane (kernel row $w_0,w_1,w_2$) of a 10-pixel row. Output $o_c$ uses pixels $c$, $c{+}1$ and $c{+}2$ ($o_c=x_c w_0+x_{c+1}w_1+x_{c+2}w_2$; $o_0$ and $o_6$ shown), giving 8 outputs. (a) Kept pixels (blue) are the first three and all nonzeros. (b) Each kept pixel becomes a record $\{last,z,x\}$; the offset $z$ counts the zeros skipped before it, so record $m$ sits at column $c=m+z$. A window is three consecutive records.}
\label{fig:example}

\vspace{1pt}
\begin{tikzpicture}[font=\scriptsize, >={Latex[length=1.1mm]},
  blk/.style={draw, rounded corners=1pt, fill=blue!5, align=center, inner sep=1pt},
  reg/.style={draw, fill=orange!18, minimum width=5.6mm, minimum height=4.4mm, inner sep=0pt},
  dly/.style={draw, fill=gray!15, minimum width=3.2mm, minimum height=3.2mm, inner sep=0pt, font=\tiny},
  add/.style={draw, circle, fill=white, inner sep=0pt, minimum size=2.7mm, font=\tiny},
  lbl/.style={font=\fontsize{5.5}{6}\selectfont, text=black!75}]
\node[blk, minimum height=21mm, minimum width=8.5mm] (win) at (0.45,-0.8) {line\\buffers\\+\\3$\times$3\\window};
\foreach \j/\y in {0/0,1/-0.8,2/-1.6} {
  \node[blk, minimum width=9.5mm, minimum height=5mm] (pe\j) at (1.5,\y) {PE$_{\j}$\\[-1pt]$x\,w_{\j}$};
  \draw[->] (0.875,\y) -- (pe\j.west);
}
\node[dly] (d1) at (2.35,-0.8) {D};
\node[dly] (d2a) at (2.35,-1.6) {D}; \node[dly] (d2b) at (2.85,-1.6) {D};
\draw (pe0.east) -- (3.05,0); \draw[->] (pe1.east) -- (d1); \draw (d1) -- (3.05,-0.8);
\draw[->] (pe2.east) -- (d2a); \draw[->] (d2a) -- (d2b); \draw (d2b) -- (3.05,-1.6);
\node[lbl] at (3.0,0.14) {$p_0$}; \node[lbl] at (3.0,-0.66) {$p_1$}; \node[lbl] at (3.25,-1.46) {$p_2$};
\node[lbl] at (2.45,0.33) {align taps};
\node[reg] (R2) at (5.6,0) {$R_2$};
\draw[->] (3.05,0) -- (R2.west);
\node[add] (a1) at (4.0,-0.72) {$+$};
\draw[->] (4.0,-0.45) node[above, lbl, inner sep=0.5pt] {$R_2$} -- (a1.north);
\draw[->] (3.05,-0.8) -- (3.3,-0.8) |- (a1.west);
\draw[->] (3.3,-0.8) |- (4.65,-0.92);
\draw[->] (a1.east) -- (4.65,-0.72);
\filldraw[fill=green!12] (4.65,-0.6) -- (4.95,-0.7) -- (4.95,-0.94) -- (4.65,-1.04) -- cycle;
\node[reg] (R1) at (5.6,-0.82) {$R_1$};
\draw[->] (4.95,-0.82) -- (R1.west);
\node[add] (a01) at (4.0,-1.42) {$+$};
\node[add] (a02) at (4.0,-1.88) {$+$};
\draw[->] (4.0,-1.18) node[above, lbl, inner sep=0.5pt] {$R_1$} -- (a01.north);
\draw[->] (4.0,-2.13) node[below, lbl, inner sep=0.5pt] {$R_2$} -- (a02.south);
\draw (3.05,-1.6) -- (3.45,-1.6);
\draw[->] (3.45,-1.6) |- (a01.west); \draw[->] (3.45,-1.6) |- (a02.west);
\draw[->] (3.45,-1.6) -- (3.45,-1.65) -- (4.65,-1.65);
\draw[->] (a01.east) -- (4.65,-1.42); \draw[->] (a02.east) -- (4.65,-1.88);
\filldraw[fill=green!12] (4.65,-1.28) -- (4.95,-1.4) -- (4.95,-1.9) -- (4.65,-2.02) -- cycle;
\node[reg] (R0) at (5.6,-1.65) {$R_0$};
\draw[->] (4.95,-1.65) -- (R0.west);
\node[lbl] at (5.6,0.31) {col $c$};
\node[lbl] at (5.6,-0.49) {col $c{-}1$};
\node[lbl] at (5.6,-1.32) {col $c{-}2$};
\tikzset{ml/.style={font=\fontsize{5}{5.5}\selectfont, inner sep=0.4pt, anchor=south east, text=red!55!black}}
\node[ml] at (4.6,-0.69) {1}; \node[ml, anchor=north east] at (4.6,-0.95) {${\ge}2$};
\node[ml] at (4.6,-1.39) {1}; \node[ml] at (4.6,-1.62) {${\ge}3$}; \node[ml, anchor=north east] at (4.6,-1.91) {2};
\node[blk] (det) at (2.35,-2.62) {jump $\Delta$:\ $c'{=}c{+}1\Rightarrow1$\\[-1pt]$c'{=}c{+}2\Rightarrow2$, else $\ge3$};
\draw[->] (win.south) |- node[pos=0.3, left, lbl] {$c'$} (det.west);
\draw[->, dashed] (det.east) -| node[pos=0.8, right, lbl] {$\Delta$} (4.8,-1.97);
\draw[->, dashed] (4.8,-2.4) -- (5.1,-2.4) |- (4.95,-1.1) -- (4.8,-1.1) -- (4.8,-0.99);
\node[blk, minimum width=11mm, minimum height=21mm] (wr) at (6.85,-0.8) {retire\\$\min(\Delta,3)$\\columns\\[1pt]$\downarrow$\\write\\register};
\draw[->] (R2.east) -- (R2.east -| wr.west); \draw[->] (R1.east) -- (R1.east -| wr.west); \draw[->] (R0.east) -- (R0.east -| wr.west);
\node[blk, minimum width=14mm] (bk) at (6.85,-3.4) {$B$ BRAM banks\\[-1pt]write once,\\[-1pt]read + clear};
\draw[->] (wr.south) -- (bk.north);
\node[blk] (sum) at (4.45,-3.4) {$\Sigma$ 3 lanes\\$B$ col/cycle};
\draw[->] (bk.west) -- (sum.east);
\node[blk, fill=gray!12] (l3) at (2.35,-3.4) {6:1 channel adder\\[-1pt]bias, ReLU, quant};
\draw[->] (sum.west) -- (l3.east);
\end{tikzpicture}

\caption{One lane of an engine (three lanes per engine, one per kernel row). Tap $j$ is delayed by $j$ lane steps so that all products of a record arrive together. Red labels: the jump $\Delta$ (Sec.~\ref{sec:arch}) that selects each mux input; finished columns are written once into $B$ banks. The engine-wide stall (ce) is not shown.}
\label{fig:arch}

\vspace{3pt}
{\scriptsize\setlength{\tabcolsep}{2pt}
\begin{tabular}{@{}cccllll@{}}
\toprule
rec. & col & jump & written & \multicolumn{3}{c}{partial sums in the window after the record}\\
\cmidrule(l){5-7}
$m$ & $c$ & $\Delta$ & (once) & $R_0$: col $c{-}2$ & $R_1$: col $c{-}1$ & $R_2$: col $c$\\
\midrule
0 & 0 & -- & -- & $\cdot$ & $\cdot$ & $o_0$: $5w_0$\\
1 & 1 & 1 & -- & $\cdot$ & $o_0$: $5w_0{+}0w_1$ & $o_1$: $0w_0$\\
2 & 2 & 1 & -- & $o_0$: $5w_0{+}0w_1{+}3w_2$ & $o_1$: $0w_0{+}3w_1$ & $o_2$: $3w_0$\\
3 & 6 & 4 & $o_0,o_1,o_2$ & $o_4$: $7w_2$ & $o_5$: $7w_1$ & $o_6$: $7w_0$\\
4 & 7 & 1 & $o_4$ & $o_5$: $7w_1{+}2w_2$ & $o_6$: $7w_0{+}2w_1$ & $o_7$: $2w_0$\\
5 & 9 & 2 & $o_5,o_6$ & $o_7$: $2w_0{+}4w_2$ & $\cdot$ & $\cdot$\\
end & & & $o_7$ & & & \\
\bottomrule
\end{tabular}}
\caption{The example of Fig.~\ref{fig:example} in the lane accumulator after tap alignment: $o_0$ is complete at record 2 and written once when record 3 jumps to column 6. $\cdot$ = outside the row; $o_3$ is never written and reads as zero.}
\label{fig:trace}
\end{figure}

\section{Row-Local Accumulator}
\label{sec:arch}
Fig.~\ref{fig:arch} shows one lane of an engine. Six engines feed the 6:1 adder (Fig.~\ref{fig:flow}): for conv1 the one input channel is broadcast to all six, each with its own filter; for conv2 each engine gets its own input channel and the adder sums them for one filter.

\textbf{Lanes and tap alignment.} PE $p$ belongs to lane $\lfloor p/3\rfloor$ and picks one of the three pixels of its own kernel row. In window $w$ the three PEs of a lane compute $x_w w_0$, $x_{w+1} w_1$ and $x_{w+2} w_2$ in the same cycle. With gaps these go to different outputs (Sec.~\ref{sec:scatter}). We therefore delay the product of tap $j$ by $j$ steps of the lane (the D registers in Fig.~\ref{fig:arch}). After this delay, each step brings the three products of \emph{one record}, $x w_0$, $x w_1$ and $x w_2$, which always go to columns $c$, $c{-}1$ and $c{-}2$ (bounded reach).

\textbf{Three-register window.} Each lane keeps three registers $R_0, R_1, R_2$ per filter, for columns $c{-}2$, $c{-}1$ and $c$ of the latest record. The next record arrives at column $c'$ with products $p_0, p_1, p_2$; each product carries its output column, so $c'$ is the column of $p_0$. Its \emph{jump} is $\Delta = c' - c$ (the same role as the shift step in \cite{liu21}), and the registers are updated as follows ($R_2 \leftarrow p_0$ in every case):
\begin{center}\footnotesize\setlength{\tabcolsep}{5pt}
\begin{tabular}{@{}cccl@{}}
\toprule
$\Delta$ & $R_0 \leftarrow$ & $R_1 \leftarrow$ & retired (written)\\
\midrule
1 & $R_1 + p_2$ & $R_2 + p_1$ & $R_0$\\
2 & $R_2 + p_2$ & $p_1$ & $R_0, R_1$\\
$\ge 3$ & $p_2$ & $p_1$ & $R_0, R_1, R_2$\\
\bottomrule
\end{tabular}
\end{center}
\textbf{Retirement rule:} the registers that shift out are complete (in-order completion) and are written to memory, $\min(\Delta,3)$ per record and oldest first; at the end of a row, all registers still holding a column retire. \textbf{Finding $\Delta$.} The new column is compared with stored $c{+}1$ and $c{+}2$ (equal: $\Delta{=}1$ or 2; else $\Delta{\ge}3$); both possible sums are formed first, keeping the adder off the compare path.

\textbf{Write once, without bank conflicts.} Each kernel-row lane keeps unfinished sums in its three registers. When a sum can no longer receive another product (\textbf{in-order completion}), the lane writes it to memory exactly once. Each lane has its own buffer, divided into $B$ banks, and the columns of a row are interleaved across the banks: output column $a$ is stored in bank $a\bmod B$ at address $\lfloor a/B\rfloor$. A lane can finish at most three consecutive columns in one cycle, so $B=3$ banks are enough to write them together; we use $B=4$ unless stated, because bank and address are then simple bit fields of $a$. E.g., columns 6, 7, 8 go to banks 2, 3, 0.

Each bank has two halves, selected by a bit $h$. While one half receives the current output row, the other supplies the previous row for readout. A row of up to $W_{\max}$ columns needs $\lceil W_{\max}/B\rceil$ addresses per bank; its memory address is $\{h,\lfloor a/B\rfloor\}$. The halves swap after each row; the engine stalls (\emph{ce}$=$0), with all registers holding, only if it needs a half that is still being read. The readout clears locations for reuse, so columns that received no contribution produce zero.

Once a row is ready, readout visits address $k=0,1,\ldots,\lceil W/B\rceil-1$ in all $B$ banks of all three lanes. For each column, the three lane values are added to form the engine's output, producing up to $B$ pixels per cycle. Aligned windows and separate lane buffers let the lanes be combined here, without the inter-row RMW of \cite{liu21}.

\subsection{Packing two filters}
\label{sec:pack}
Zero weights waste multipliers: in Liu et al.\ \cite{liu21,liu23} a zero weight still occupies one, and \cite{liu23} skips either activation or weight zeros, whichever are sparser. RSNN \cite{rsnn} also merges sparse kernels, but over time: each PE takes their nonzero weights one per cycle, skips only weight zeros, and adds each product to a partial-sum buffer by RMW. We instead pack two filters in space, on a window's free multipliers, skipping both kinds of zeros in one pass. Each PE has a weight, a pixel select and a 1-bit filter tag. A PE freed by a zero weight can compute the other filter if it stays in its own kernel row (Table~\ref{tab:pack}). Each lane keeps a separate set of window registers per filter, and both sets shift together because the columns depend only on the pixels. Both filters are written in one cycle, as two fields of a word, and read out one after the other while the engine fills the other half; it waits only if a row fills faster than both are read. The 6:1 adder sums each filter over the six input channels, so all engines in a pass use the same two filters. Each kernel row can hold at most three nonzero weights of the two filters together; other pairs run in separate passes.

\begin{table}[t]
\centering\footnotesize
\caption{Packing two filters into one engine}
\label{tab:pack}
\setlength{\tabcolsep}{3pt}
\begin{tabular*}{\columnwidth}{@{\extracolsep{\fill}}cclccc@{}}
\toprule
filter A & filter B & kernel row & PE & weights & tags\\
\midrule
$a_0\,a_1\,0$ & $b_0\,0\,0$ & 0 & 0,1,2 & $a_0,a_1,b_0$ & 0,0,1\\
$a_3\,a_4\,0$ & $0\,b_4\,0$ & 1 & 3,4,5 & $a_3,a_4,b_4$ & 0,0,1\\
$a_6\,a_7\,0$ & $0\,0\,b_8$ & 2 & 6,7,8 & $a_6,a_7,b_8$ & 0,0,1\\
\bottomrule
\end{tabular*}\\[2pt]
\raggedright\scriptsize A is zero in its right column, B off its diagonal; PEs keep their row, a tag picks the filter.

\vspace{3pt}
\caption{Accumulator ablation at 128$\times$128, one engine (cycles)}
\label{tab:ablation}
\setlength{\tabcolsep}{1.6pt}
\begin{tabular*}{\columnwidth}{@{\extracolsep{\fill}}lrrrrrr@{}}
\toprule
$s$ & 0 & 0.25 & 0.51 & 0.69 & 0.80 & 0.90\\
\midrule
frontend bound & 17,022 & 13,570 & 9,550 & 6,470 & 4,605 & 2,967\\
updates / output (earlier) & 1.00 & 2.93 & 2.69 & 2.05 & 1.44 & 0.81\\
earlier engine (queue) & 48,402 & 140,407 & 128,806 & 98,020 & 69,326 & 39,040\\
SCNN-style, 4 banks & 17,055 & 22,159 & 19,394 & 14,923 & 10,901 & 6,636\\
SCNN-style, 18 banks$^\ast$ & 17,055 & 13,851 & 10,527 & 8,219 & 6,095 & 4,688\\
\textbf{this work (B=4)} & \textbf{17,055} & \textbf{13,603} & \textbf{9,583} & \textbf{6,503} & \textbf{4,866} & \textbf{4,603}\\
this work / frontend & 1.002 & 1.002 & 1.003 & 1.005 & 1.06 & 1.55\\
\% of ideal speedup & -- & 94 & 88 & 80 & 69 & 37\\
\bottomrule
\end{tabular*}\\[2pt]
\raggedright\scriptsize Up to $s{=}0.69$ this work is 33 cycles above the frontend bound: pipeline drain, as the last row's readout (126 columns, 4 per cycle: 32+1 cycles) has no next row to hide it. SCNN-style: our implementation of a crossbar into banked accumulators \cite{scnn} on the same frontend, with same-column merging and an \emph{ideal} single-cycle RMW; $^\ast$18 banks for nine multipliers (2 per multiplier; \cite{kim22} found 1.5 most efficient). A real 2-cycle BRAM RMW would add 12--38\% to the 18-bank cycles at $s{=}0.25$--$0.8$.

\end{table}

\begin{table*}[t]
\centering
\caption{Activation-sparse accelerators: platform, workloads and how partial sums are merged}
\label{tab:cmp}
\scriptsize\setlength{\tabcolsep}{2.8pt}
\begin{tabular}{@{}llllllllll@{}}
\toprule
Design & Platform & Bits & Sparsity & MACs / DSP & MHz & Workloads & Merging & Conflicts & Extra cost\\
\midrule
Sparse-PE \cite{sparsepe} & Z-7100$^\dagger$ & 8b & W+A & 9 (252) / -- & 200 & VGG, MobileNet & gather: chunk adder & none & selector\\
SCNN \cite{scnn} & ASIC (sim.) & 16b & W+A & 1024 / -- & -- & AlexNet, VGG & crossbar into banks & yes & RMW per update\\
SparTen \cite{sparten} & ASIC (sim.) & -- & W+A & -- & -- & -- & gather: match pairs & none & index matching\\
NullHop \cite{nullhop} & Z-7100 & 16b & A & 128 / 128 & 60 & VGG16 & shift 1 col./cycle & none & 1 MAC per kernel\\
Liu \cite{liu21,liu23} & 55\,nm (syn.) & 8b & A or W & 144 / -- & 100--200 & VGG, MobileNet & registers per kernel row & none & row RMW \cite{liu21}\\
GAAS \cite{gaas} & 65\,nm (sim.) & -- & W+A & 256 PEs / -- & 200 & VGG, MobileNet & hashed, assoc.\ banks & reduced & reordering\\
Zhang \cite{zhang25} & ZCU102 & 8b & W+A & 1024 / 528 & 200 & VGG16, ResNet-50 & hashed into PE array & 1.1\% left & reordering, caches\\
Kim \cite{kim22} & 180\,nm (syn.) & -- & W+A & 1024 / -- & 89 & AlexNet, SqueezeNet & crossbar into banks & reduced & bigger crossbar\\
Pham \cite{pham25} & ZU+$^\ddagger$ & 8b & W+A & 1024 (64) / 0 & 100 & ResNet, MobileNet & 3 cached output rows & none & crossbar, row cache\\
PASS \cite{pass} & ZCU102 & 16b & A & 1024 / 1024 & 200 & VGG16, ResNet-18 & adder tree & none & crossbar, +50\% LUT\\
\textbf{SkipLane} & Z-7020 (board) & 8b & A (+W packing) & 54 / 54 & 140 & LeNet-5 & 3-register lane & none & 3 banked row buffers\\
\bottomrule
\end{tabular}\\[2pt]
\raggedright\scriptsize --: not reported. Throughput is not compared, as platforms and workloads differ. A/W: activations/weights. Built: $^\dagger$9-multiplier core, $^\ddagger$64-MAC cluster; rest simulated.
\end{table*}

\section{Evaluation}
\label{sec:eval}
\textbf{Setup.} The earlier, new and SCNN-style engines are simulated in Icarus Verilog with the same frontend. Every output is compared bit for bit with a NumPy model, for images from 8$\times$8 to 224$\times$224, one and six engines, one and two packed filters, and through the 6:1 adder. Activations are unsigned 8-bit, weights signed 8-bit, sums 32-bit. Zeros are placed at random unless stated, and cycles are counted from the first input to the last output. The full design (Table~\ref{tab:impl}) is implemented in Vivado 2023.2 on the xc7z020-1, out of context. The engines close at 140\,MHz both alone and in the full board system, where the scheduler and input DMA run at 200\,MHz; only the output DMA write-back runs at 125\,MHz, behind a stream clock converter.

\begin{figure}[t]
\centering
\begin{tikzpicture}
\begin{axis}[width=0.98\columnwidth, height=27mm, xlabel={activation sparsity $s$}, ylabel={speedup vs.\ own dense},
  xmin=-0.02, xmax=0.92, ymin=0, ymax=6.4, ytick={0,2,4,6}, legend style={font=\scriptsize, at={(0.5,1.02)}, anchor=south, draw=none, /tikz/every even column/.append style={column sep=6pt}}, legend columns=2,
  tick label style={font=\scriptsize}, label style={font=\scriptsize}, grid=major, grid style={gray!20}]
\addplot[black, dashed, domain=0:0.83, samples=40] {1/(1-x)};
\addlegendentry{ideal $1/(1-s)$}
\addplot[blue!70!black, thick, mark=*, mark size=1.4pt] coordinates {(0,1.00)(0.252,1.25)(0.506,1.78)(0.695,2.62)(0.804,3.50)(0.901,3.71)};
\addlegendentry{this work, $B{=}4$}
\addplot[blue!70!black, mark=o, mark size=1.6pt, only marks] coordinates {(0.901,5.65)};
\addlegendentry{this work, $B{=}8$}
\addplot[orange!80!black, thick, mark=square*, mark size=1.3pt] coordinates {(0,1.00)(0.252,0.345)(0.506,0.376)(0.695,0.494)(0.804,0.698)(0.901,1.24)};
\addlegendentry{earlier engine (queue)}
\end{axis}
\end{tikzpicture}
\caption{Speedup over own dense input (128$\times$128, one engine, RTL sim.).}
\label{fig:speedup}
\end{figure}
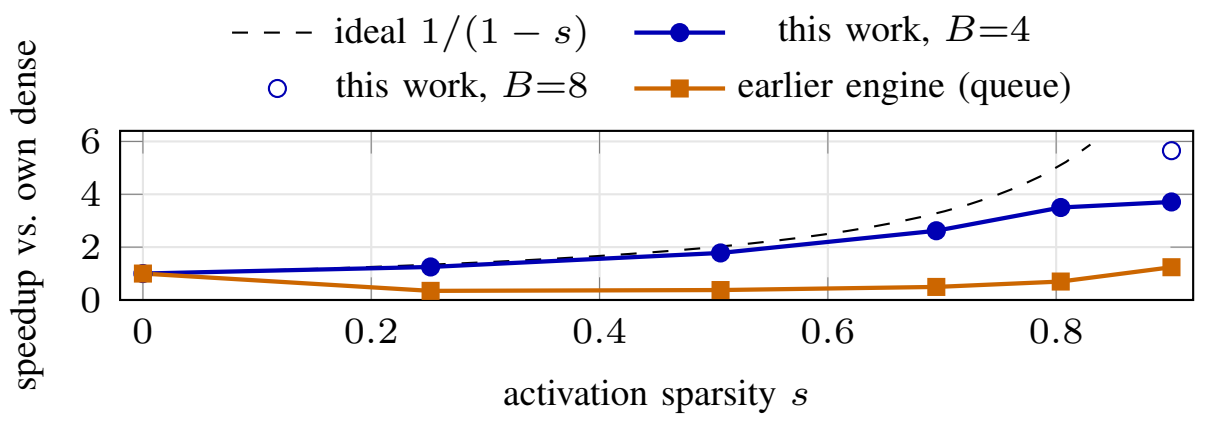

\textbf{Speedup.} The \emph{frontend} (line buffers, window generator and PEs) forms one window per cycle, so it sets a lower bound on cycles (Table~\ref{tab:ablation}). Up to 69\% sparsity the new engine is within 0.5\% of this bound (Fig.~\ref{fig:speedup}). This is 80--94\% of the ideal speedup and 10--15$\times$ the speed of the earlier engine (Table~\ref{tab:ablation}), with 23\% fewer LUTs but 50\% more FFs and 81 vs.\ 69 BRAM36 (Table~\ref{tab:impl}): the banked lane buffers spend memory to remove accumulation stalls. Against an SCNN-style scatter accumulator on the same frontend
(Table~\ref{tab:ablation}), ours is 1.10--1.26$\times$ faster at 50--80\% sparsity even when the baseline gets 18 banks and an ideal RMW, and 2.0--2.3$\times$ faster against a four-bank scatter baseline (ours has four banks per lane, 12 per engine). Above 80\% sparsity the 4-column readout limits speed; $B{=}8$ raises the speedup at 90\% from 3.7$\times$ to 5.7$\times$.

\textbf{Packing.} With 89\% zero weights and 80\% zero activations, two filters packed into one pass run 6.9$\times$ faster than two separate dense passes at 128$\times$128 with $B{=}8$: about 3.5$\times$ from skipping zeros and 2$\times$ from sharing the pass. The 1-bit tag limits the weight gain to 2$\times$.

\begin{table}[t]
\centering\footnotesize
\caption{Implementation on xc7z020 (six engines)}
\label{tab:impl}
\setlength{\tabcolsep}{2pt}
\begin{tabular*}{\columnwidth}{@{\extracolsep{\fill}}lrrrr@{}}
\toprule
 & LUT & FF & BRAM36 & DSP\\
\midrule
dense engine \cite{corevision} & 3,890 & 5,263 & 6 & 54\\
earlier sparse (synth.) & 27,270 & 12,678 & 69 & 54\\
\textbf{this work (routed)} & \textbf{21,082} & \textbf{19,013} & \textbf{81} & \textbf{54}\\
\midrule
$f_{max}$ / power & \multicolumn{4}{l}{140\,MHz (alone and in system) / 0.61\,W (est.)}\\
\bottomrule
\end{tabular*}

\vspace{4pt}
\caption{LeNet-5 (3$\times$3 kernels) on the PYNQ-Z2, 100 MNIST test digits}
\label{tab:lenet}
\setlength{\tabcolsep}{3pt}
\begin{tabular}{@{}lrrrrrr@{}}
\toprule
 & \multicolumn{3}{c}{engine$^\ast$ (cycles)} & \multicolumn{3}{c}{end to end (cycles / $\mu$s)}\\
\cmidrule(lr){2-4}\cmidrule(l){5-7}
 & dense & sparse & speedup & dense & sparse & speedup\\
\midrule
conv1 & 824 & 425 & 1.94$\times$ & 824 / 5.49 & 612 / 4.37 & 1.26$\times$\\
conv2/pass & 195 & 186 & 1.05$\times$ & 195 / 1.30 & 208 / 1.49 & 0.87$\times$\\
\bottomrule
\end{tabular}\\[2pt]
\raggedright\scriptsize Board counters, means over 100 digits (all 1,700 sparse passes bit-exact). $^\ast$Input buffered; sparse ranges 327--559, 146--216. End to end: input streamed, first DMA beat to last output; time at board clocks (sparse 140\,MHz, dense 150\,MHz). The dense engine takes one pixel per cycle, so its engine and end-to-end cycles are equal.
\end{table}

\textbf{LeNet-5: speed versus throughput.} The dense engine peaks at 16.2\,GOPS; the sparse design reaches 25.0 and 36.9 dense-equivalent GOPS at 51\% and 69\% sparsity on 128-pixel rows. Table~\ref{tab:lenet} gives board results (82\% zero pixels). Conv1 needs 1.94$\times$ fewer engine cycles. End to end, conv1 is 1.26$\times$ faster: the input sets this time, as the scheduler examines every pixel, zeros included, at one per cycle (200\,MHz). Conv2 inputs are 71\% zero, but rows are only 13 pixels wide and each pass waits for its least sparse channel, so conv2 gains only 1.05$\times$ in cycles and is 0.87$\times$ the dense speed end to end. The sparse engine therefore suits highly sparse layers with few channels and wide rows, where multipliers, not area, are the limit; the smaller dense engine suits many channels.

\textbf{Comparison.} In Table~\ref{tab:cmp}, \emph{scatter} designs \cite{scnn,gaas,zhang25,kim22} can send two products to one bank in the same cycle; such conflicts are reduced, not removed (51\% to 1.1\% in \cite{zhang25}, with 48k LUTs and 253k FFs of caches). On AlexNet, an SCNN-style design reaches only 45\% of its conflict-free speed; 76\% of its conflicts are products for the same output, which no bank mapping avoids \cite{kim22}. \emph{Gather} designs \cite{sparsepe,sparten,stitchx,kwon} need matching or selection logic. PASS \cite{pass} skips zeros inside dense windows and sums each window's nonzero products in an adder tree: nothing is merged, but each window still takes a cycle, so sparsity saves DSPs, not cycles. Liu et al.\ \cite{liu21} use the same in-order property, shifting a kernel row's partial sums through three registers by the jump between nonzero pixels. But their kernel rows process an input row one after another, each adding to the previous one's results by RMW of a buffer; our lanes work on the same output row in the same pass, write their banks once, and are added at readout. \cite{liu23} works on 3$\times$9 tiles but skips only one kind of zero and takes channels one at a time. ORSAS \cite{orsas} and Pham et al.\ \cite{pham25} also compute one output row at a time, but add each product by RMW into a row buffer only 19 and 14 columns wide. Our lanes cover full rows and hold each column in registers until complete; finished columns go to different banks, so banks never conflict and each output is written once and read once, with no read-back. The price is a fixed dataflow (3$\times$3, stride 1; weight zeros only via packing).

\section{Conclusion}
In SkipLane, write-once lanes with conflict-free banks keep the engine at its frontend bound up to 70\% sparsity and beat an SCNN-style accumulator, and filter packing puts multipliers idled by zero weights to work. Six engines run LeNet-5 bit-exact on a PYNQ-Z2 at 140\,MHz. Next are a multi-pixel scheduler and VGG-style layers.

\section*{Acknowledgment}
Claude (Anthropic) assisted with some of the RTL, proofreading and language editing; the authors reviewed all content.

\end{document}